\documentclass[conference,twocolumn,twoside,10pt]{IEEEtran}
\usepackage[utf8]{inputenc}
\usepackage{amsmath}
\usepackage{cite}
\usepackage{amssymb}
\usepackage{stackrel}
\usepackage{booktabs}
\usepackage{mathtools}
\usepackage{amsthm}
\usepackage{comment}
\usepackage{bm}
\usepackage{enumerate}
\usepackage{enumitem}
\usepackage[nolist]{acronym}
\usepackage{pgfplots}
\usepackage{microtype}
\usepackage{mathrsfs}
\usepackage{sansmath}
\usepackage{bbm}
\usepackage{amsfonts}
\pgfplotsset{compat=newest}
\pgfplotsset{plot coordinates/math parser=false}
\usepackage{tabularx}
\usepackage{float}
\usepackage{makecell}
\usepackage{collcell}
\usepackage{tikz}
\usepackage{setspace}
\usepackage{xcolor}
\usepackage{multirow}
\usepackage{graphicx}
\usetikzlibrary{decorations.pathreplacing,calc}
\usetikzlibrary{plotmarks}
\usetikzlibrary{matrix}
\usepackage[ruled,linesnumbered]{algorithm2e}
\usepackage{comment}
\ifCLASSOPTIONcompsoc
    \usepackage[caption=false, font=normalsize, labelfont=sf, textfont=sf]{subfig}
\else
\usepackage[caption=false, font=footnotesize]{subfig}
\fi

\title{A Grant-free Coded Random Access Scheme for Near-field Communications}

\author{\IEEEauthorblockN{
Enrico Testi\IEEEauthorrefmark{1}\IEEEauthorrefmark{3},
Giulia Torcolacci\IEEEauthorrefmark{1}\IEEEauthorrefmark{3},
Nicol\`o Decarli\IEEEauthorrefmark{2}\IEEEauthorrefmark{3},
Davide Dardari\IEEEauthorrefmark{1}\IEEEauthorrefmark{3},
Enrico Paolini\IEEEauthorrefmark{1}\IEEEauthorrefmark{3}
}\\
\IEEEauthorblockA{\IEEEauthorrefmark{1} DEI, Universit\`a di Bologna, 40136 Bologna, Italy}
\IEEEauthorblockA{\IEEEauthorrefmark{2} IEIIT, National Research Council (CNR), 40136 Bologna, Italy}
\IEEEauthorblockA{\IEEEauthorrefmark{3} National Laboratory of Wireless Communications (WiLab), CNIT, 40136 Bologna, Italy}
}

\definecolor{carrotorange}{rgb}{0.93, 0.57, 0.13}

\acrodef{$P_{EM}$}{probability of emulation, or false alarm}
\acrodef{$P_{FA}$}{probability of false alarm}
\acrodef{$P_{MD}$}{probability of missed detection}
\acrodef{$P_{D}$}{probability of detection}

\acrodef{3D}{three-dimensional}
\acrodef{5G}{5th Generation}
\acrodef{6G}{6th Generation}
\acrodef{ACF}{autocorrelation function}
\acrodef{ACG}{automatic	gain control}
\acrodef{ACI}{adjacent channel interference}
\acrodef{ACK}{acknowledge}
\acrodef{AcR}{autocorrelation receiver}
\acrodef{ADC}{analog-to-digital converter}
\acrodef{AF}{amplify \& forward}
\acrodef{AFL}{anchor-free localization}
\acrodef{AGNSS}{assisted-GNSS}
\acrodef{AGPS}{assisted GPS}
\acrodef{AI}{artificial intelligence}
\acrodef{AIC}{Akaike information criterion}
\acrodef{AO}{alternating optimization}
\acrodef{AOA}{angle-of-arrival}
\acrodef{AOD}{angle-of-departure}
\acrodef{AOT}{approximate optimum threshold}
\acrodef{AP}{access point}
\acrodef{API}{application programming interface}
\acrodef{ASK}{amplitude shift keying}
\acrodef{ASNR}{accumulated signal-to-noise ratio}
\acrodef{AUB}{asymptotic union bound}
\acrodef{AWGN}{additive white Gaussian noise}
\acrodef{BAN}{body area network}
\acrodef{BAV}{balanced antipodal Vivaldi}
\acrodef{BCH}{Bose Chaudhuri Hocquenghem}
\acrodef{BEP}{bit error probability}
\acrodef{BER}{bit error rate}
\acrodef{BF}{brute force}
\acrodef{BFC}{block fading channel}
\acrodef{BIC}{Bayesian information criterion}
\acrodef{BLUE}{best linear unbiased estimator}
\acrodef{BPAM}{binary pulse amplitude modulation}
\acrodef{BPF}{bandpass filter}
\acrodef{BPPM}{binary pulse position modulation}
\acrodef{bps}{bits per second}
\acrodef{BPSK}{binary phase shift keying}
\acrodef{BPZF}{band-pass zonal filter}
\acrodef{BS}{base station}
\acrodef{BSC}{binary symmetric channel}
\acrodef{BTB}{Bellini-Tartara bound}
\acrodef{c.c.d.f.}{complementary cumulative distribution function}
\acrodef{c.d.f.}{cumulative distribution function}
\acrodef{CAD}{computer-aided design}
\acrodef{CAIC}{consistent Akaike information criterion}
\acrodef{CAP}{continuous aperture phased}
\acrodef{CCF}{cross correlation function}
\acrodef{CCI}{co-channel interference}
\acrodef{CD}{cooperative diversity}
\acrodef{CDMA}{code division multiple access}
\acrodef{CEOT}{channel ensemble optimum threshold}
\acrodef{CEP}{codeword error probability}
\acrodef{CFAR}{constant	 false alarm rate}
\acrodef{ch.f.}{characteristic function}
\acrodef{CH}{cluster head}
\acrodef{CIR}{channel impulse response}
\acrodef{CL}{centroid localization}
\acrodef{CM}{channel model}
\acrodef{CNR}{clutter-to-noise ratio}
\acrodef{CP}{ciclic prefix}
\acrodef{CPR}{channel pulse response}
\acrodef{CR}{channel response}
\acrodef{CRB}{Cram\'{e}r-Rao bound}
\acrodef{CRC}{cyclic redundancy check}
\acrodef{CRA}{coded random access}
\acrodef{CREMA}{coded random electromagnetic access}
\acrodef{CRLB}{Cram\'{e}r-Rao lower bound}
\acrodef{CS}{clock skew}
\acrodef{CSCG}{circularly symmetric complex Gaussian}
\acrodef{CSI}{channel state information}
\acrodef{CSMA}{carrier sense multiple access}
\acrodef{CSRA}{coded spatial random access}
\acrodef{SE-CSRA}{single-element CSRA}
\acrodef{CSS}{chirp spread spectrum}
\acrodef{CTS}{clear-to-send}
\acrodef{CW}{continuous wave}
\acrodef{DAA}{detect and avoid}
\acrodef{DAB}{digital audio broadcasting}
\acrodef{DBB}{digital base band}
\acrodef{DBPSK}{differential binary phase shift keying}
\acrodef{DCM}{dual-carrier modulation}
\acrodef{DDP}{detected direct path}
\acrodef{DF}{detect \& forward}
\acrodef{DFT}{Discrete Fourier transform}
\acrodef{DFMS}{monopole dual feed stripline antenna}
\acrodef{DGPS}{differential GPS}
\acrodef{DLL}{delay-locked loop}
\acrodef{DNN}{deep neural network}
\acrodef{DoD}{Department of Defense}
\acrodef{DoF}{degrees of freedom}
\acrodef{DP}{direct path}
\acrodef{DR}{detection rate}
\acrodef{DRT}{distance ratio test}
\acrodef{DS-SS}{direct-sequence spread-spectrum}
\acrodef{DS}{delay spread}
\acrodef{DTR}{differential transmitted-reference}
\acrodef{DTT}{Diffraction Tomography Theory}
\acrodef{DVB-H}{digital video broadcasting\,--\,handheld}
\acrodef{DVB-T}{digital video broadcasting\,--\,terrestrial}
\acrodef{e.m.}{electromagnetic}
\acrodef{ECC}{European Community Commission}
\acrodef{ED}{energy detection}
\acrodef{EDR}{energy detector receiver}
\acrodef{EFIM}{equivalent Fisher information matrix}
\acrodef{EIRP}{effective radiated isotropic power}
\acrodef{EKF}{extended Kalman filter}
\acrodef{KKT}{Karush–Kuhn–Tucker}
\acrodef{ELAA}{extremely large aperture array}
\acrodef{ELP}{equivalent low-pass}
\acrodef{EM}{electromagnetic}
\acrodef{EMCB}{extended Miller Chang bound}
\acrodef{EME}{minimum eigenvalue ratio detector}
\acrodef{EMI}{electromagnetic interference}
\acrodef{ENP}{estimated noise power}
\acrodef{ESA}{European Space Agency}
\acrodef{EU}{European Union}
\acrodef{EVD}{eigenvalue decomposition}
\acrodef{FAR}{false alarm rate}
\acrodef{FCC}{Federal Communications Commission}
\acrodef{FDMA}{frequency division multiple access}
\acrodef{FDMA}{frequency division multiple access}
\acrodef{FEC}{forward error correction}
\acrodef{FEC}{forward error correction}
\acrodef{FFD}{full function device}
\acrodef{FFR}{full function reader}
\acrodef{FF}{far-field}
\acrodef{FFT}{fast Fourier transform}
\acrodef{FG}{factor graph}
\acrodef{FH-SS}{frequency-hopping spread-spectrum}
\acrodef{FH}{frequency-hopping}
\acrodef{FIM}{Fisher information matrix}
\acrodef{FLL}{Frequency-locked loop}
\acrodef{FS}{frame synchronization}
\acrodef{FT}{Fourier Transform}
\acrodef{GA}{Gaussian approximation}
\acrodef{GA-CSRA}{genie-aided CSRA}
\acrodef{GD}{gradient descent}
\acrodef{GDOP}{geometric dilution of precision}
\acrodef{GLR}{generalized likelihood ratio}
\acrodef{GLRT}{generalized likelihood ratio test}
\acrodef{GML}{generalized maximum likelihood}
\acrodef{GPRS}{general packet radio service}
\acrodef{GPS}{global positioning system}
\acrodef{HAP}{high altitude platform}
\acrodef{HCRB}{hybrid Cram\'{e}r-Rao bound}
\acrodef{HDSA}{high-definition situation-aware}
\acrodef{Hi-RADIAL}{High-accuracy RAdio Detection, Identification, And Localization}
\acrodef{HMM}{hidden Markov model}
\acrodef{HPA}{high-power amplifier}
\acrodef{HPBW}{half power beam width}
\acrodef{HW}{hardware}
\acrodef{i.i.d.}{independent, identically distributed}
\acrodef{ICT}{information and communication technologies}
\acrodef{IE}{informative element}
\acrodef{IEEE}{Institute of Electrical and Electronics Engineers}
\acrodef{IF}{intermediate frequency}
\acrodef{IFFT}{inverse fast Fourier transform}
\acrodef{IMF}{ideal matched filter}
\acrodef{IMU}{inertial measurement unit}
\acrodef{INR}{interference-to-noise ratio}
\acrodef{INS}{inertial navigation system}
\acrodef{IoT}{Internet of things}
\acrodef{IIoT}{industrial Internet of things}
\acrodef{INS}{inertial navigation system}
\acrodef{IR-UWB}{impulse radio UWB}
\acrodef{IR}{impulse radio}
\acrodef{IRI}{inter-reader interference}
\acrodef{IRS}{intelligent reflecting surface} 
\acrodef{ISAC}{integrated sensing and communications}
\acrodef{ISI}{inter-symbol interference} 
\acrodef{isi}{intra-symbol interference} 
\acrodef{ISM}{industrial, scientific and medical}
\acrodef{ISNR}{interference-plus-signal-to-noise-ratio}
\acrodef{ISP}{inverse scattering problem}
\acrodef{IT}{interference temperature}
\acrodef{ITC}{information theoretic criteria}
\acrodef{JBSF}{jump back and search forward}
\acrodef{JF}{just forward}
\acrodef{KF}{Kalman filter}
\acrodef{KKT}{Karush–Kuhn–Tucker}
\acrodef{LDC}{low duty cycle}
\acrodef{LDMA}{location division multiple access}
\acrodef{LDPC}{low density parity check}
\acrodef{LEO}{localization error outage}
\acrodef{LG}{Laguerre-Gaussian}
\acrodef{LIS}{large intelligent surface}
\acrodef{LLR}{log-likelihood ratio}
\acrodef{LLRT}{log-likelihood ratio test}
\acrodef{LRT}{likelihood ratio test}
\acrodef{LNA}{low-noise amplifier}
\acrodef{LOS}{line-of-sight}
\acrodef{LRT}{likelihood ratio test}
\acrodef{LS}{least square}
\acrodef{LS}{least squares}
\acrodef{M-PSK}{$M$-ary phase shift keying}
\acrodef{M-QAM}{$M$-ary quadrature amplitude modulation}
\acrodef{m.g.f.}{moment generating function}
\acrodef{MAC}{medium access control}
\acrodef{MAE}{mean absolute error}
\acrodef{MAI}{multiple access interference}
\acrodef{MAN}{metropolitan area network}
\acrodef{MAP}{maximum a posteriori}
\acrodef{MB-OFDM}{multi-band OFDM}
\acrodef{MB-UWB}{multi-band UWB}
\acrodef{MB}{multi-band}
\acrodef{MC}{multi-carrier}
\acrodef{MCB}{Miller Chang bound}
\acrodef{MCRB}{modified Cram\'{e}r-Rao bound}
\acrodef{MDD}{minimum distance distribution}
\acrodef{MDL}{minimum description length}
\acrodef{MF}{matched filter}
\acrodef{MGF}{moment generating function}
\acrodef{MI}{mutual information}
\acrodef{MIMO}{multiple-input multiple-output}
\acrodef{MISO}{multiple-input single-output}
\acrodef{ML}{maximum likelihood}
\acrodef{MM}{min-max}
\acrodef{MMA}{massive multiple access}
\acrodef{MME}{maximum-minimum eigenvalue ratio detector}
\acrodef{mMIMO}{massive MIMO}
\acrodef{mm-waves}{Millimeter Waves}
\acrodef{MMSE}{minimum mean-square error}
\acrodef{MPC}{multipath component}
\acrodef{MRC}{maximal ratio combining}
\acrodef{MTC}{machine-type communication}
\acrodef{MS}{mobile station}
\acrodef{MSB}{most significant bit}
\acrodef{MSE}{mean squared error}
\acrodef{NMSE}{normalized mean squared error}
\acrodef{MSK}{minimum shift keying}
\acrodef{MUI}{multi-user interference}
\acrodef{MUR}{multistatic radar}
\acrodef{MVU}{minimum variance unbiased}
\acrodef{MZZB}{modified Ziv-Zakai bound}
\acrodef{NB}{narrowband}
\acrodef{NBI}{narrowband interference}
\acrodef{NEO}{navigation error outage}
\acrodef{NFER}{near-Þeld electromagnetic ranging}
\acrodef{NF}{near-field}
\acrodef{NFF}{near-field focused}
\acrodef{NL}{nonlinear}
\acrodef{NLOS}{non-line-of-sight}
\acrodef{NP}{Neyman-Pearson}
\acrodef{NSD}{normalized standard deviation}
\acrodef{NTIA}{National Telecommunications and Information Administration}
\acrodef{NTP}{network time protocol}
\acrodef{NU}{network user}
\acrodef{OAM}{orbital angular momentum} 
\acrodef{OC}{optimum combining}
\acrodef{OFDM}{orthogonal frequency division multiplexing}
\acrodef{OOK}{on-off keying}
\acrodef{OP}{outage probability}
\acrodef{OT}{optimum threshold}
\acrodef{P-Max}{$P$-Max}  %suggestion, use with \acl{P-Max}
\acrodef{p.d.f.}{probability density function}
\acrodef{p.m.f.}{probability mass function}
\acrodef{PA}{power amplifier}
\acrodef{PAM}{pulse amplitude modulation}
\acrodef{PAN}{personal area network}
\acrodef{PAR}{peak-to-average ratio}
\acrodef{P-CRLB}{Posterior Cramer-Rao Lower Bound}
\acrodef{PCA}{principal component analysis}
\acrodef{PD}{probability of detection}
\acrodef{PDP}{power delay profile}
\acrodef{PE}{probability of emulation}
\acrodef{PEB}{position error bound}
\acrodef{PEC}{perfect electric conductor}
\acrodef{PEP}{packet error probability}
\acrodef{PF}{particle filter}
\acrodef{PFA}{probability of false alarm}
\acrodef{PHY}{physical}
\acrodef{PL}{path-loss}
\acrodef{PLL}{phase-locked loop}
\acrodef{PLR}{packet loss rate}
\acrodef{PMD}{probability of missed detection}
\acrodef{PN}{pseudo-noise}
\acrodef{PSF}{point spread function}
\acrodef{ppm}{part-per-million}
\acrodef{PPM}{pulse position modulation}
\acrodef{PR}{pseudo-random}
\acrodef{PRake}{partial rake}
\acrodef{PRF}{pulse repetition frequency}
\acrodef{PRP}{pulse repetition period}
\acrodef{PSD}{power spectral density}
\acrodef{PSEP}{pairwise synchronization error probability}
\acrodef{PSNR}{peak signal to noise ratio}
\acrodef{PSK}{phase shift keying}
\acrodef{PSVD}{product singular value decomposition}
\acrodef{PSWF}{prolate spheroidal wave function}
\acrodef{PU}{primary user}
\acrodef{QAM}{quadrature amplitude modulation}
\acrodef{QoS}{quality of service}
\acrodef{QPSK}{quadrature phase shift keying}
\acrodef{R.V.}{random variable}
\acrodef{RADAR}{radar}
\acrodef{RCS}{radar cross section}
\acrodef{RDL}{"random data limit"}
\acrodef{REM}{radio environment map}
\acrodef{REO}{ranging error outage}
\acrodef{RF}{radio-frequency}
\acrodef{RFID}{radio-frequency identification}
\acrodef{RFR}{reduced function reader}
\acrodef{RFT}{reduced function tag}
\acrodef{RII}{ranging information intensity}
\acrodef{RIS}{reconfigurable intelligent surface}
\acrodef{rms}{root mean square}
\acrodef{RMSE}{root-mean-square error}
\acrodef{ROC}{receiver operating characteristic}
\acrodef{ROI}{region of interest}
\acrodef{RRC}{root raised cosine}
\acrodef{RSN}{radar sensor network}
\acrodef{RSS}{received signal strength}
\acrodef{RSSI}{received signal strength indicator}
\acrodef{RTLS}{real time locating systems}
\acrodef{RTT}{round-trip time}
\acrodef{S-V}{Saleh-Valenzuela}
\acrodef{SA}{simulated annealing}
\acrodef{SaG}{stop-and-go}
\acrodef{SAR}{synthetic aperture radar}
\acrodef{SBS}{serial backward search}
\acrodef{SBSMC}{serial backward search for multiple clusters}
\acrodef{SCM}{supply chain management}
\acrodef{SCR}{signal-to-clutter ratio}
\acrodef{SE}{spectral efficiency}
\acrodef{SEP}{symbol error probability}
\acrodef{SIC}{successive interference cancellation}
\acrodef{SIS}{small intelligent surface}
\acrodef{SFD}{start frame delimiter}
\acrodef{SIC}{successive interference cancellation}
\acrodef{SIMO}{single-input multiple-output}
\acrodef{SINR}{signal-to-interference plus noise ratio}
\acrodef{SIR}{signal-to-interference ratio}
\acrodef{SISO}{single-input single-output}
\acrodef{SNR}{signal-to-noise ratio}
\acrodef{SoC}{system on chip}
\acrodef{SoO}{signal of opportunity}
\acrodef{SoP}{system on package}
\acrodef{SOT}{sub-optimum threshold}
\acrodef{SPAWN}{sum-product algorithm over a wireless network}
\acrodef{SPEB}{squared position error bound}
\acrodef{SPMF}{single-path matched filter}
\acrodef{SQNR}{signal-to-quantization-noise ratio}
\acrodef{SRE}{smart radio environment}
\acrodef{SS}{spread spectrum}
\acrodef{ST}{simple thresholding}
\acrodef{SU}{secondary user}
\acrodef{SVD}{singular value decomposition}
\acrodef{SW}{software}
\acrodef{SW}{sync word}
\acrodef{TDE}{time delay estimation}
\acrodef{TDL}{tapped delay line}
\acrodef{TDMA}{time division multiple access}
\acrodef{TDOA}{time difference-of-arrival}
\acrodef{TH}{time-hopping}
\acrodef{THz}{TeraHertz}
\acrodef{TNR}{threshold-to-noise ratio}
\acrodef{TOA}{Time-of-arrival}
\acrodef{TOF}{time-of-flight}
\acrodef{TPC}{transmit power control}
\acrodef{TR}{transmitted-reference}
\acrodef{TS}{tabu search}
\acrodef{TSVD}{truncated singular value decomposition}
\acrodef{TV}{total variation denoising}
\acrodef{UAV}{unmanned aerial vehicle}
\acrodef{UB}{union bound}
\acrodef{UCA}{uniform circular array}
\acrodef{UDP}{undetected direct path}
\acrodef{UE}{User Equipment}
\acrodef{UHF}{ultra-high frequency}
\acrodef{ULA}{uniform linear array}
\acrodef{ULP}{user location protocol}
\acrodef{UMP}{uniformly most powerful}
\acrodef{UMPI}{uniformly most powerful invariant}
\acrodef{URA}{uniform rectangular array}
\acrodef{UT}{user terminal}
\acrodef{UTC}{coordinated universal time}
\acrodef{UTM}{universal transverse Mercator}
\acrodef{UTRA}{UMTS terrestrial radio access}
\acrodef{UAV}{unmanned aerial vehicle}
\acrodef{UUV}{unmanned underwater vehicle}
\acrodef{UWB}{ultrawide-band}
\acrodef{UWBcap}[UWB]{Ultrawide band}
\acrodef{VFIL}{virtual force iterative localization}
\acrodef{VGA}{variable-gain amplifier}
\acrodef{VNA}{vector network analyzer}
\acrodef{WAF}{wall attenuation factor}
\acrodef{WB}{wideband}
\acrodef{WBI}{wideband interference}
\acrodef{WCL}{weighted centroid localization}
\acrodef{WED}{wall extra delay}
\acrodef{WiMAX} {worldwide interoperability for microwave access}
\acrodef{WLAN}{wireless local area network}
\acrodef{WLS}{weighted least squares}
\acrodef{WMAN}{wireless metropolitan area network}
\acrodef{WPAN}{wireless personal area networks}
\acrodef{WRAPI}{wireless research application programming interface}
\acrodef{WSN}{wireless sensor network}
\acrodef{WSR}{wireless sensor radar}
\acrodef{WSS}{wide-sense stationary}
\acrodef{WWB}{Weiss-Weinstein bound}
\acrodef{WWLB}{Weiss-Weinstein lower bound}
\acrodef{ZF}{zero forcing}
\acrodef{ZZB}{Ziv-Zakai bound}
\acrodef{ZZLB}{Ziv-Zakai lower bound}
\acrodef{XL-MIMO}{extremely large-scale multiple-input multiple-output}
\definecolor{carrotorange}{rgb}{0.93, 0.57, 0.13}
\definecolor{lightblue}{RGB}{10, 106, 164}
\newcommand{\rec}{\bm{y}}
\newcommand{\Rec}{\bm{Y}}

\newcommand{\m}{\mathsf{W}}

\newcommand{\z}{\bm{z}}

\renewcommand{\P}{\mathcal{P}}

\newcommand{\List}{\mathcal{L}}
\newcommand{\C}{\mathcal{C}}

\newcommand{\Pt}{\rho}

\newcommand{\PL}{P_{\mathrm{L}}}

\let\oldnl\nl% Store \nl in \oldnl
\newcommand{\nonl}{\renewcommand{\nl}{\let\nl\oldnl}}% Remove line number for one line

\newcommand{\cn}{{\mathcal{CN}}} %Complex Gaussian RV

\newcommand{\boldb}{{\bf b}}

\newcommand{\boldh}{{\bf h}}

\newcommand{\boldp}{{\bf p}}

\newcommand{\bolds}{{\bf s}}

\newcommand{\boldw}{{\bf w}}
\newcommand{\boldx}{{\bf x}}
\newcommand{\boldy}{{\bf y}}

\newcommand{\boldH}{{\bf H}}
\newcommand{\boldI}{{\bf I}}

\newcommand{\boldW}{{\bf W}}

\newcommand{\boldY}{{\bf Y}}

\newcommand{\Lt}{L_{\text{T}}}
\newcommand{\Lr}{L_{\text{R}}}

\renewcommand{\Pt}{P_{\text{T}}}

\newcommand{\Nt}{N_{\text{T}}}
\newcommand{\Nr}{N_{\text{R}}}
\newcommand{\Np}{N_{\text{P}}}
\newcommand{\Nd}{N_{\text{D}}}
\newcommand{\Ns}{N_{\text{S}}}

\newcommand{\hkrtc}{{h}_{r, t}^{(k)}} 

\begin{document}
\bstctlcite{IEEEexample:BSTcontrol}
\maketitle

\begin{abstract}
The \ac{IIoT} is revolutionizing industrial processes by facilitating massive machine-type communications among countless interconnected devices. To efficiently handle the resulting large-scale and sporadic traffic, grant-free random access protocols—especially \ac{CRA}—have emerged as scalable and reliable solutions. At the same time, advancements in wireless hardware, including extremely large-scale MIMO arrays and high-frequency communication (e.g., mmWave, Terahertz), are pushing network operations into the near-field propagation regime, allowing for dense connectivity and enhanced spatial multiplexing. This paper proposes an innovative approach that combines near-field spatial multiplexing with the interference mitigation capabilities of \ac{CRA}, utilizing an extremely large aperture array at the access point. This integration improves reliability and reduces access latency, offering a robust framework for \ac{IIoT} connectivity in next-generation 6G networks.
\end{abstract}
\acresetall

\section{Introduction}\label{sec:intro}

The \ac{IIoT} is revolutionizing industries by enabling interconnected devices and systems that facilitate real-time monitoring, automation, and data-driven decision-making, serving as the backbone of Industry 4.0\cite{DawSaaGho:J17,Bockelmann2016:Massive}. In the 6G era, these \ac{MTC} systems will demand unprecedented scalability, improved battery efficiency, and diverse reliability and latency requirements \cite{Pokhrel2020:Towards}. Meeting these requirements necessitates innovations in \ac{MAC} and \ac{PHY} layers, particularly to support grant-free \ac{MMA} protocols, which are well-suited for managing the dense and sporadic traffic characteristic of \ac{MTC}.
Grant-free access allows devices to transmit without prior scheduling, reducing the signaling overhead and latency but risking collisions when many devices communicate simultaneously. Recent developments in \ac{CRA} techniques, especially those leveraging \ac{SIC}, have enhanced reliability by allowing devices to send replicas over multiple resources, significantly boosting the chances of successful decoding \cite{TesTraPao:J24,paolini2015:magazine}.
When high carrier frequencies such as mmWave or Terahertz are adopted in conjunction with extremely large scale (XL) MIMO arrays, propagation occurs within the radiative near-field region, where devices can leverage spatial multiplexing even when operating in \ac{LOS} to enhance their communication capability \cite{DarDec:J21,XLMIMOtut:J24a,XLMIMOtut:J24b}. In this setup, multi-antenna \ac{IIoT} devices can establish parallel channels (namely, communication modes), thereby enhancing network efficiency. However, to exploit high carrier frequencies in \ac{MTC}, proper multi-antenna beamforming gain at both the transmitter and receiver is required to counteract the increased path loss and exploit the spatial multiplexing capabilities, thus requiring precise channel estimation, which may be impractical due to overhead constraints \cite{Cav:J23}.
Recent studies have explored leveraging the near-field region to enhance communication performance.
In \cite{SunetAl:J15}, the authors propose a beam division multiple access (BDMA) scheme for massive MIMO, organizing users into non-overlapping beams and minimizing channel estimation overhead. 
In \cite{WuDai:J23}, \ac{LDMA} for near-field \ac{MIMO} is introduced to improve spectrum efficiency by leveraging beam focusing.
In \cite{cui2022near}, the authors address beam training in near-field communications, focusing on \ac{CSI} acquisition.
A complex bilinear inference algorithm for XL-MIMO activity detection is proposed in \cite{iimori2022grant}.
In \cite{de2023qos}, a QoS-aware joint user scheduling and power allocation technique for XL-MIMO systems is proposed.
A beamspace modulation strategy for XL-MIMO communications that enhances spectral efficiency through spatial degrees of freedom by utilizing few radio frequency chains is introduced in \cite{guo2023beamspace}. 
Despite their contributions, all these approaches rely heavily on complex channel estimation processes, which are impractical for the sporadic traffic typical of \ac{MTC} in \ac{IIoT} networks.

To address this gap, in this paper we propose a novel near-field communication strategy that combines grant-free \ac{MMA} with multi-modal communication in the near-field. This is performed by utilizing an \ac{ELAA} at the \ac{AP}, so that multi-antenna \ac{IIoT} devices can transmit across both spatial and temporal domains without extensive channel estimation procedures. By clustering received signals at the \ac{ELAA} to exploit the combining gain and applying \ac{SIC}, this strategy mitigates collisions and increases the overall throughput, supporting dense, uncoordinated \ac{IIoT} networks with high reliability and efficiency.

\section{System Model}\label{sec:systemmodel}

\begin{figure}
    \centering
    \includegraphics[width=0.99\linewidth]{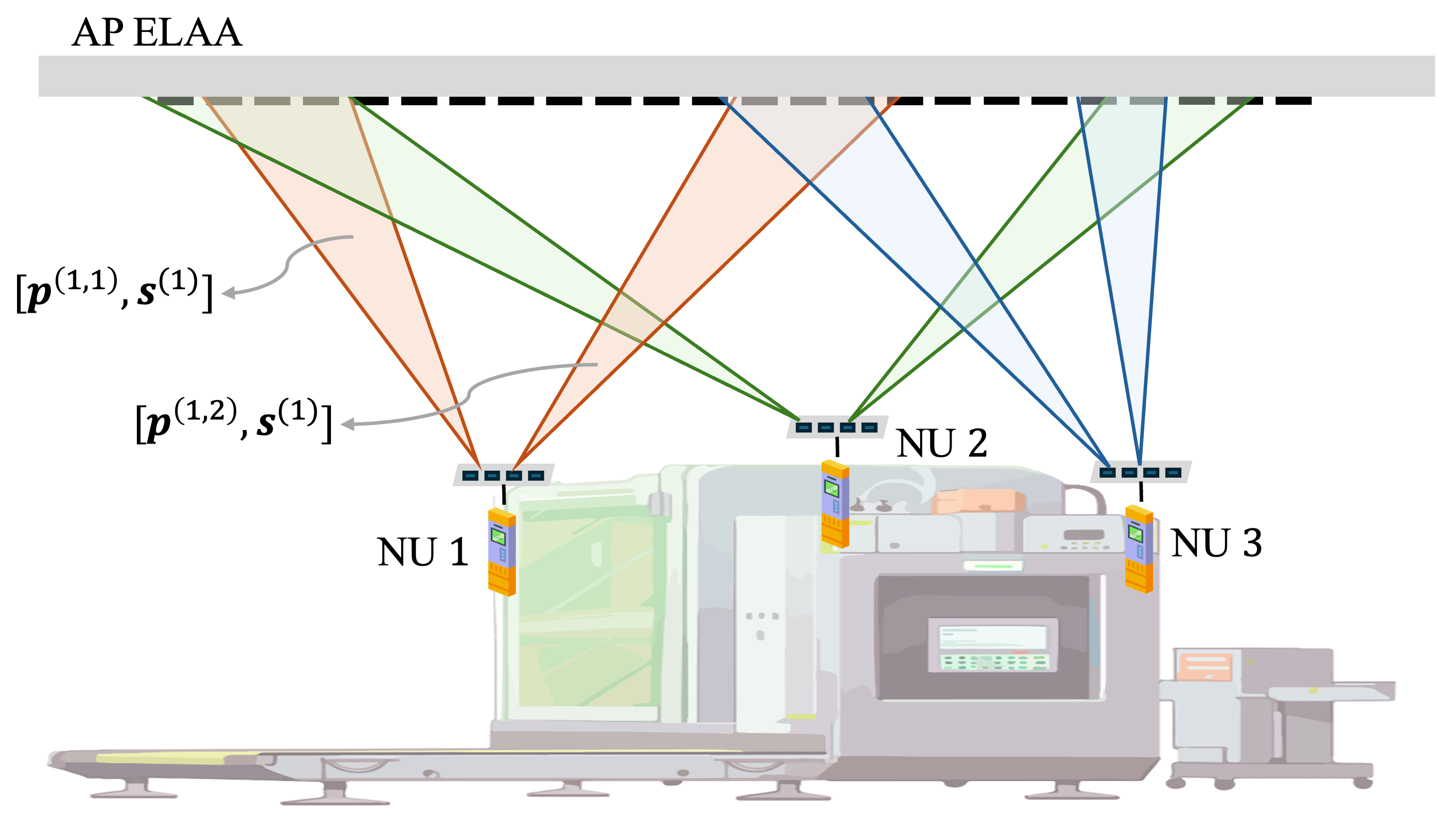}
    \caption{IIoT scenario with three multi-antenna NUs, placed on an industrial machine, transmitting their signals towards an overhead AP equipped with an ELAA, according to the coded spatial random access scheme.}
    \label{fig:scenario}
\end{figure}

Let us consider an industrial setting, such as a manufacturing line in a factory as illustrated in Fig.~\ref{fig:scenario}. In this scenario, $N$ \acp{NU}, i.e., machine-type devices, randomly transmit signals towards the ceiling, where an \ac{ELAA} is installed and acts as receiving \ac{AP}. Specifically, the \ac{ELAA} could be realized as a linear \ac{LIS} or a so-called \textit{radio stripe} \cite{ShaBjoLar:C20}, which we assume to be deployed along the $y$-axis. Each of the transmitting \acp{NU} is equipped with a \ac{ULA} having $N_{\mathrm{T}}$ antenna elements located at $\mathbf{p}_{\mathrm{T}, i}^{(k)}=[x_{\mathrm{T}, i}^{(k)}, y_{\mathrm{T}, i}^{(k)}, z_{\mathrm{T}, i}^{(k)}]^T$, with $i=1,2, \ldots, N_{\mathrm{T}}$ and $k =1,2,\ldots, N$. Let us denote by 
%$\Lt = \Lt^{(k)} \forall k \in \mathcal{K}$$, 
$\Lt$ the length of the \ac{NU}'s array, which we assume to be identical for all users.
All \acp{NU} are considered to be located in the $xy$-plane, positioned parallel to the receiving \ac{AP}, with no misalignment with respect to the $y$-axis between the arrays. Such a condition can be realized with proper deployment of the \ac{NU} antennas along the machine. Regardless of the implementation strategy for the \ac{ELAA} at the \ac{AP}, we model it as a long \ac{ULA} of length $\Lr$ encompassing $\Nr$ antenna elements, each of them located at $\mathbf{p}_{\mathrm{R}, r}=\left[x_{\mathrm{R}, r}, y_{\mathrm{R}, r}, z_{\mathrm{R}, r}\right]^T$, $r=1,2, \ldots, \Nr$. Moreover, we have $\Nt\ll\Nr$ and all antenna elements, both at the \ac{AP} and the \acp{NU}, are uniformly spaced with an inter-element distance of $\lambda/2$, where $\lambda$ denotes the wavelength. 

Let us assume that the generic (say the $k$-th) \ac{NU}, if active, transmits a sequence of $\Ns$ complex symbols ${\boldx^{(k)}= [ x_{1}^{(k)}, x_{2}^{(k)}, \ldots, x_{\Ns}^{(k)}] \in \mathbb{C}^{1 \times \Ns}}$,  with $||\boldx^{(k)}||^2 = \Ns$. We denote by $\Pt$ the \acp{NU}' transmit power.

At the \ac{AP} side, when all $N$ users are active, the received signal $\boldY = [ \boldy_1, \boldy_2, \ldots, \boldy_{\Nr}]^T\in \mathbb{C}^{\Nr \times \Ns}$ is expressed as
\begin{align}\label{eq:rx_sig}
 \boldY =  \sqrt{\Pt} \sum_{k=1}^{N}\boldH^{(k)} \,\boldb^{(k)} \otimes \boldx^{(k)} + \boldW 
\end{align}
where $\boldy_r$ represents the signal received at the $r$-th \ac{AP} array element, $\boldH^{(k)}=\{ \hkrtc \}\in \mathbb{C}^{\Nr \times \Nt}$ denotes the complex channel matrix associated with the $k$-th \ac{NU}, $\boldb^{(k)} = [b_1^{(k)}, b_2^{(k)}, \ldots, b_{\Nt}^{(k)}]^T \in \mathbb{C}^{\Nt \times 1}$ is the beamforming vector of the $k$-th \ac{NU}, $\boldW = [\boldw_1, \boldw_2, \ldots, \boldw_{\Nr}]^T\in \mathbb{C}^{\Nr\times \Ns}\,$ is a matrix of \ac{AWGN} samples, where $\boldw_{r} \in \mathbb{C}^{\Ns \times 1}$ and $\,\boldw_{r} \sim \cn (\mathbf{0}, \sigma^2\boldI_{\Ns \times \Ns})$, and $t$ and $r$ indicate the $t$-th element of the $k$-th \ac{NU}'s array and the $r$-th element of the \ac{AP} \ac{ELAA}, respectively. The noise variance is $\sigma^2=B k_B T_0 \mathrm{NF}$, where $B$ is the bandwidth, $k_B$ is the Boltzmann constant, $T_0$ is the equivalent noise temperature, and $\mathrm{NF}$ denotes the noise figure of the receiver. 

In \ac{LOS} free-space conditions, the elements of the channel matrix $\boldH^{(k)}$ in \eqref{eq:rx_sig} are given by
\begin{align} \label{eq:hkri}
    &\hkrtc = \frac{\lambda}{4\, \pi\, d_{t,r}^{(k)}} \, e^{- \jmath \frac{2\, \pi}{ \lambda}\,d_{t,r}^{(k)}  }.  
\end{align}
The quantity $d_{t,r}^{(k)}$ represents the distance between the $(\mathbf{p}_{\mathrm{T}, t}^{(k)}, \mathbf{p}_{\text{R}, r})$ antenna pairs. 

Given \textit{(i)} the extremely large aperture of the \ac{AP} antenna array, \textit{(ii)}  the high operating frequency employed (e.g., mm-Waves or THz), and \textit{(iii)} the short communication distances (e.g., in the same order of the \ac{AP} array aperture), the system operates in the near-field regime \cite{XLMIMOtut:J24a}.
Accordingly, in the following, we assume that all the \acp{NU} are located in the radiative near-field region of the receiving \ac{AP}, where the distance $d^{(k)}$ between the center of the \ac{ULA} of the $k$-th \ac{NU} and the center of the \ac{AP} \ac{ELAA} is lower than the Fraunhofer distance \cite{Bal:B15}, that is $d^{(k)}<2L_{\text{R}}^2/{\lambda}$.

This definition identifies the so-called Fresnel region, where the \ac{EM} wavefront can no longer be approximated as planar and its spherical curvature must be considered instead. In this near-field condition, since we assumed the presence of multiple antennas at the \ac{NU} side, it becomes possible to exploit several communication modes to communicate between each \ac{NU} and the \ac{AP} \cite{DarDec:J21}. Specifically, this means that the channel described by the matrix $\boldH^{(k)}$ has a rank larger than one even considering \ac{LOS} channel conditions and without multipath, hence allowing data communications between the \ac{NU} and the \ac{AP} along multiple parallel channels. 
It can be shown that, for a given geometry and operating wavelength under the $\Lt\ll \Lr$ (i.e., $\Nt \ll \Nr$) condition, the number of strongly coupled communication modes (i.e., with significant values of the corresponding coupling coefficients) is given by \cite{DecDar:J21} 
\begin{equation}\label{eq:Nmodes}
   R_{\text{max}}^{(k)} =  \left\lfloor 1 + \frac{2 \Lt \Lr}{\lambda \sqrt{4 \left(d^{(k)}\right)^2+\Lr^2}} \right\rfloor
\end{equation}
where $\left\lfloor \cdot \right\rfloor$ identifies the floor operator.
Therefore, these $R_{\text{max}}^{(k)}$ modes can be exploited to establish orthogonal communication channels between the $k$-th \ac{NU} and the \ac{AP}.  

\section{Coded Spatial Random Access}\label{sec:CRA}
In this section, we delineate the novel grant-free random access scheme, operating in the near field region, and the corresponding processing scheme at the \ac{AP}.

\subsection{Uplink Users' Transmission}
Optimal communication between the  $k$-th \ac{NU} and the \ac{AP} within the near-field region should be realized by exploiting simultaneously the available  $R_{\text{max}}^{(k)}$ communications modes. In this scenario, capacity maximization can be achieved by employing dedicated beamforming vectors at the \acp{NU} and corresponding combining vectors at the \ac{AP}. However, this approach imposes significant challenges, including the need for complex channel estimation, singular value decomposition (SVD) of the channel matrix, exchange of additional data (i.e., overhead), and precise synchronization between the \acp{NU} and the \ac{AP}. These stringent requirements are impractical in \ac{MTC} environments, particularly when dealing with a large number of sporadically active devices, as commonly observed in \ac{IIoT} applications. Therefore, in the following, we propose a grant-free \ac{MMA} scheme that leverages the multiple communication modes available in the near-field scenario under consideration.
Specifically, it has been shown in \cite{DecDar:J21} that the $R_{\text{max}}^{(k)}$ communication modes can be easily realized considering simple beam steering vectors at the transmitting \ac{NU}. This happens since \acp{NU} are equipped with much fewer antennas than the \ac{AP}; thus, despite \acp{NU} being within the near-field region of the \ac{AP} antenna array, it is likely that the \ac{AP} is located outside the near-field region of the transmitting \acp{NU}. In such a condition, the number of well-coupled communication modes $R_{\text{max}}^{(k)}$ corresponds to the number of orthogonal beams generated by the $k$-th \ac{NU} capable of intercepting the \ac{AP} \ac{ELAA}. Thus, it can be asserted that communication modes correspond to distinct beam steering directions from the \ac{NU} towards the \ac{AP}. These directions must be chosen ensuring orthogonality in the angular domain among the different beams \cite{DecDar:J21}. Multiple transmission directions in space will be then adopted within the grant-free \ac{MMA} scheme hereby proposed; for this reason, we refer to our \ac{CRA} approach as \ac{CSRA}.
We assume that time is slotted and, in each time slot, $K$ out of $N$ \acp{NU} sporadically and unpredictably wake up to transmit short information messages in a grant-free manner, competing to deliver one message each to the \ac{AP}. The uplink transmission operates in a slot-synchronous mode, meaning that there is a common time reference between the \acp{NU} and the \ac{AP}. This synchronization is maintained by a beacon signal transmitted by the \ac{AP} at the beginning of each slot. After waking up, an active \ac{NU} $k$ waits for the beacon sent by the \ac{AP} and competes to transmit one $m$-bit message $\m^{(k)} \in \{1,\dots,2^m\}$, within the subsequent slot. We denote by $\mathcal{K}$ the subset of \acp{NU} which are active during a generic time slot and assume that the \ac{AP} is unaware of which \ac{NU} are active during any given time slot.

The access protocol operates as follows. Let us assume that all $K$ active \acp{NU} encode their messages using the same channel encoder and map the coded bits onto the same complex constellation. Consequently, each \ac{NU} obtains a data payload $\bolds^{(k)} \in \mathbb{C}^{1 \times \Nd}$ consisting of $\Nd$ complex symbols. Then, each active \ac{NU} selects $R^{(k)} \leq R_{\text{max}}^{(k)}$ transmission directions and transmits a packet $\boldx^{(k,j)} = [\boldp^{(k,j)},\bolds^{(k)}] \in \mathbb{C}^{1 \times \Ns}$, with ${j=1,2,\dots,R^{(k)}}$, in such directions, where $\boldp^{(k,j)} \in \mathbb{C}^{1 \times \Np}$ is a pilot sequence drawn from a set $\P$, with $\left| \P\right| = P$ and $P\ll K$, of mutually orthogonal pilots. Therefore, we have $\Ns = \Np + \Nd$. We refer to the transmission of the $k$-th \ac{NU} in each of its $R^{(k)}$ selected directions as a ``replica". Moreover, each replica arriving at the \ac{AP} is assumed to be perfectly synchronized with one of the designated time slots.
Therefore, during uplink transmission, each \ac{NU} in $\mathcal{K}$ first sends the chosen pilot sequences across its $R^{(k)}$ transmission directions, followed by the payload data. The corresponding vector of signal samples received at the $r$-th \ac{AP} \ac{ELAA} antenna element is denoted by $\tilde{\boldy}_r = [\tilde{\boldy}_r^{\text{P}} , \tilde{\boldy}_r^{\text{D}} ] \in \mathbb{C}^{1 \times \Ns}$, in which $\tilde{\boldy}_r^{\text{P}} \in \mathbb{C}^{1 \times \Np}$ is the vector of samples corresponding to the pilot and is given by
\begin{align}\label{eq:rec_pilot}
    \tilde{\boldy}_r^{\text{P}} = \sqrt{\Pt} \sum_{k\in\mathcal{K}} \boldh_{r}^{(k)} \sum_{j=1}^{R^{(k)}}\boldb^{(k,j)} \otimes \boldp^{(k,j)} + \bm{\omega}_{r} 
\end{align}
where $r=1,2,\dots,\Nr$, $\boldh_{r}^{(k)}$ is the vector of complex channel gains between \ac{ELAA} antenna element $r$ and \ac{NU} $k$, i.e., the $r$-th row of the channel matrix $\boldH^{(k)}$, ${\bm{\omega}_{r} \sim \mathcal{C}\mathcal{N} (\bm{0}_{\Np}, \sigma^2 \mathbf{I}_{\Np})}$ is the vector of \ac{AWGN} noise samples, and  ${\boldb^{(k,j)} = [b_1^{(k,j)}, b_2^{(k,j)}, \ldots, b_{\Nt}^{(k,j)}] \in \mathbb{C}^{\Nt \times 1}}$ is the beamforming vector of \ac{NU} $k$ employed to transmit in the $j$-th direction, among the $R^{(k)}$ available ones. Notably, the overall beamforming vector $\boldb^{(k)}$ in \eqref{eq:rx_sig} is $\boldb^{(k)}=1/R^{(k)} \sum_{j=1}^{R^{(k)}}\boldb^{(k,j)}$.
Similarly, the received signal samples corresponding to the payload transmission, i.e., $\tilde{\boldy}_r^{\text{D}} \in \mathbb{C}^{1 \times \Nd}$, are 
\begin{align} \label{eq:rec_payload}
    \tilde{\boldy}_r^{\text{D}} = \sqrt{\Pt} \sum_{k\in\mathcal{K}} \boldh_{r}^{(k)} \sum_{j=1}^{R^{(k)}} \boldb^{(k,j)} \otimes \bolds(\m^{(k)})  + \bm{\nu}_{r}
\end{align}
where  $r=1,2,\dots,\Nr$, $\bolds(\m^{(k)})$ is the data payload of message $\m^{(k)}$ transmitted by \ac{NU} $k$, and $\bm{\nu}_{r} \sim \mathcal{C}\mathcal{N} (\bm{0}_{\Nd}, \sigma^2  \mathbf{I}_{\Nd})$ is the vector of noise samples. Specifically, we have ${\boldw_r = [\bm{\omega}_{r},\bm{\nu}_{r}] \in \mathbb{C}^{1 \times \Ns}\,,\,\boldw_r  \sim \mathcal{C}\mathcal{N} (\bm{0}_{\Ns}, \sigma^2  \mathbf{I}_{\Ns})}$, where $\boldw_r$ identifies the $r$-th row of the noise matrix $\boldW^{(k)}$.

\subsection{Selection of the Transmission Directions}
The complete set of orthogonal transmission directions that can be spanned by the \ac{NU} antenna array can be realized by considering the classical \ac{DFT}-based codebook of beamforming vectors, frequently adopted for initial access through beam sweeping \cite{YanYanHan:C10}. In fact, the $\Nt$ antennas can span $\Nt$ different orthogonal beams (i.e., $2\Lt/\lambda$) by considering as beamforming vectors 
\begin{equation} \label{eq:DFT}
    \frac{1}{\sqrt{\Nt}}
    \begin{bmatrix}
        1 \\
        e^{- \jmath \pi n \frac{2}{\Nt}} \\
        e^{- \jmath \pi 2 n \frac{2}{\Nt}} \\
        \vdots \\
        e^{- \jmath \pi (\Nt-1) n \frac{2}{\Nt}} \\
        \end{bmatrix}
\end{equation}
for $n=\pm 1, \pm 2, \ldots, \left\lfloor \frac{\Nt}{2} \right\rfloor$. The different vectors correspond to the transmission directions towards angles \cite{DecDar:J21}
\begin{equation}
    \theta_n=\arcsin{\frac{2n}{\Nt}}\, .
\end{equation}
Specifically, the \ac{DFT} matrix \eqref{eq:DFT} for $n=\pm 1, \pm 2, \ldots, \left\lfloor \frac{\Nt}{2} \right\rfloor$ represents an orthonormal basis for the space $\mathbb{C}^{\Nt}$, thus allowing to span all the space around the transmitting antenna. The characteristic of these beams is that each one points towards the null directions of all the other beams. As anticipated, the $R_{\text{max}}^{(k)}$ communication modes correspond to the subset of these beams intercepting the receiving \ac{ELAA} at the \ac{AP}. According to this strategy, all beams could be utilized if the \ac{AP} is equipped with a theoretically infinite \ac{ELAA}. For practical \ac{ELAA} dimensions, however, each \ac{NU} should select only those beams intercepting the \ac{ELAA}, based on its position, thereby realizing $R_{\text{max}}^{(k)}$ communication modes. This can be easily implemented provided that the position of \acp{NU} is fixed.

\subsection{CSRA Processing at the \ac{AP}}
This subsection illustrates the proposed \ac{CSRA} processing scheme employed at the \ac{AP}. The \ac{CSRA} scheme is logically divided into three phases: (\textit{i}) an initial processing step that forms clusters of antenna elements based on the detected received signals; (\textit{ii}) an initial decoding attempt for all identified clusters; (\textit{iii}) a \ac{SIC} process. The \ac{CSRA} scheme is as follows.

\subsubsection{Energy detection and clustering}
Firstly, the \ac{AP} performs \ac{ED} on the channel estimates obtained from the pilot signal samples received at each antenna element. The pilot signal in \eqref{eq:rec_pilot} received at antenna element $r$ is projected along pilot $\boldp_j \in \P$, $j=1,\ldots,P$, and \ac{ML} channel estimation is obtained as
\begingroup
\allowdisplaybreaks
\begin{align}\label{eq:MLchEst}
    \hat{h}_{r}(\boldp_j) &= \frac{\boldy_r^{\text{P}} \boldp_j^H}{\sqrt{\Pt}\lVert \boldp_j \rVert^2}\,
\end{align}
\endgroup 
for all $r=1,\ldots,\Nr$.
Subsequently, \ac{ED} is performed to detect if an \ac{NU} is transmitting pilot $\boldp_j$ towards the $r$-th antenna element as
\begin{align}\label{eq:ED}
    \left|\hat{h}_{r}(\boldp_j)\right|^2 \stackrel[\mathcal{D}_0]{\mathcal{D}_1}{\gtrless} \eta 
\end{align}
where $\mathcal{D}_0$ represents the event that no signal is detected, while $\mathcal{D}_1$ is the event that a signal corresponding to the pilot $\boldp_j$ is detected at the $r$-th antenna element, and $\eta$ is the detection threshold.
The operation in \eqref{eq:ED} is performed for all pilots $\boldp_j \in \P$ with $j=1,\ldots,P$, at each individual antenna element $r$, with $r={1,2,\dots,\Nr}$. 
Once the activity status of each pilot is detected at each antenna element, a clustering operation is conducted to identify the subset of antenna elements where the signal from the same \ac{NU} is impinging. Let $\mathcal{C}_i (\mathbf{p}_j)=\{ t_1,\ldots, t_{C_{i,j}}\}$ denote the $i$-th cluster, which is composed of a set of contiguous antenna elements, indexed as $t_1,\dots,t_{C_{i,j}}$, that have received pilot $\mathbf{p}_j$ with significant power—i.e., where the corresponding estimated channel coefficient exceeds the threshold in \eqref{eq:ED}. We also denote by  $|\C_i(\mathbf{p}_j)| = C_{i,j}$ the cardinality of this set. For example, $\mathcal{C}_1 (\mathbf{p}_1)$ represents the first group of contiguous antenna elements that received pilot $\mathbf{p}_1$ with significant power. Continuing in ascending order of the antenna indices, the next group of contiguous elements that also receive significant power from pilot $\mathbf{p}_1$ forms cluster $\mathcal{C}_2 (\mathbf{p}_1)$. The outcome of this operation is a set $\mathcal{N}_{\mathrm{C}}(\mathbf{p}_j)$ of clusters for each available pilot $\mathbf{p}_j \in \mathcal{P}$, where each set comprises multiple clusters of contiguous antenna elements.

\subsubsection{Cluster-by-cluster decoding}
For each identified cluster, the \ac{AP} performs signal detection and decoding based on the linear combination of the received payload signals from all the antenna elements within the cluster. 
Adopting the conventional \ac{MRC}, the decoding of a user's message is attempted for cluster $\C_i(\boldp_j)$ starting from vector $\z_i(\boldp_j)$ obtained as
\begin{align}\label{eq:MRCsingleAP}
    \z_i(\boldp_j) &= \frac{\hat{\bm{h}}_{i}^H(\boldp_j) \Rec_{i}^{\text{D}}}{\lVert \hat{\bm{h}}_{i}(\boldp_j) \rVert^2}   \, , \, i \in \C_i(\boldp_j)
\end{align}
where $\hat{\bm{h}}_{i}(\boldp_j)=[\hat{{h}}_{t_1}(\boldp_j),\ldots,\hat{{h}}_{t_{C_{i,j}}}(\boldp_j)]^T$ is the vector of channel gains estimated over the $j$-th pilot at all the $C_{i,j}$ antenna elements in cluster $\C_i(\boldp_j)$, and $\Rec_{i}^{\text{D}}=[ (\rec_{t_1}^{\text{D}})^T,\dots,(\rec_{t_{C_{i,j}}}^{\text{D}})^T]^T$ is the matrix of aggregated received payload symbols at all the antenna elements $i \in \C_i(\boldp_j)$ composing the $i$-th cluster. If the message, sent by an \ac{NU} towards the processed cluster of antenna elements using the $j$-th pilot, is retrieved upon successful demapping and decoding performed on $\z_i(\boldp_j)$, the message is pushed to a list $\List$. This operation is first performed for all pilots $\boldp_j \in \P$ within a fixed cluster. Afterward, the same process is repeated for each cluster $i \in \mathcal{N}_{\mathrm{C}}$.

\subsubsection{\ac{SIC}}
The \ac{AP} processes the messages in the list $\List$ sequentially. 
Let us denote a generic message in the list as $\m$, and by $\mathcal{U}_{\m}$ the subset of \ac{AP} antenna elements that have received the replicas of the message $\m$ with significant power.
The \ac{AP} selects the first message in the list $\List$, i.e., $\m_{1}$, and performs the subtraction of the interference generated by the transmitted replicas of the decoded message, followed by a new decoding attempt, to all the antenna elements in which the replicas were transmitted.
The \ac{AP} can extract all of this information directly from the received message \cite{Paolini22:Irregular}. Additionally, by knowing the position of the \ac{NU}, the \ac{AP} can identify the indices of the antenna elements towards which the message replicas have been directed, i.e., $\mathcal{U}_{\m}$.
In particular, for each antenna element $u \in \mathcal{U}_{\m}$, the \ac{AP} first re-estimates the channel using the known payload, $\bolds(\m)$, as
\begin{align}\label{eq:MLchEstPay}
    \tilde{h}_{u}^{(\m)} &= \frac{\rec_{u}^{\text{D}} \bolds(\m)^H}{\lVert \bolds(\m)\rVert^2}
\end{align}
and then subtracts $\tilde{h}_{u}^{(\m)}[\boldp(\m), \bolds(\m)]$ from all the received pilot and payload samples $[\rec_{u}^{\text{P}},\rec_{u}^{\text{D}}]$.
After canceling the interference of all replicas of the messages in $\List$, the \ac{AP} performs again energy detection in each of the previously processed antenna elements, constructs new clusters of antenna elements as in step $1$, and attempts decoding of new messages by performing the operations in \eqref{eq:MLchEst} and \eqref{eq:MRCsingleAP} for each of the identified clusters. When a new message is decoded, it is pushed into the bottom of $\List$, and then \ac{SIC} is executed again. This process is repeated for every entry of $\List$ until no further messages can be decoded.

\section{Numerical Results}\label{sec:Results}
In this section, we evaluate the performance of the proposed \ac{CSRA}  scheme in an \ac{IIoT} scenario resembling a factory line, where multiple sensors installed on machinery transmit data to an \ac{ELAA} placed on the rooftop, as illustrated in Fig.~\ref{fig:scenario}. The evaluation is conducted through extensive simulations, comparing the proposed \ac{CSRA} scheme with two alternatives: \textit{(i)} a \ac{CSRA} variant without \ac{SIC}, and \textit{(ii)} a variant in which clustering is omitted (i.e., single-element processing is performed at the ELAA).
In fact, due to the grant-free nature of the communication, the \ac{AP} cannot perform a-priori channel estimation or acquire information about the active nodes, making comparisons with grant-based schemes in the literature unsuitable. Therefore, as a performance benchmark, we include a single-element variant of the \ac{CSRA} scheme, labeled CSRA-SE, where the \ac{AP} performs detection and decoding on each antenna element individually, without our proposed clustering operation that enables linear combining of received signals, thus beamforming gain at the receiver side.

\subsection{Simulation Parameters}
The system operates at a carrier frequency of $60\,$GHz. The \ac{AP} features an \ac{ELAA} positioned on the rooftop along the $y$-axis, centered at $[0,0,h_\text{AP}]$, where $h_\text{AP}=8\,$m represents the height of the factory. To minimize edge effects during the evaluation of processing schemes, the \ac{ELAA} is configured as a $20\,$m long radio stripe.
The \acp{NU} are each equipped with a linear antenna array of $\Nt=20$ elements and are positioned randomly along a segment spanning from $[0,-3\,\textrm{m},0]$ to $[0,3\,\textrm{m},0]$ on the $y$-axis.
Each \acp{NU} transmits with power $\Pt=0.1\,$mW, while $B=100\,$kHz, $T_0=290\,$K, and $\mathrm{NF}=10\,$dB. Antennas with $0\,$dBi gain are considered, and an overall implementation loss of $10\,$dB.
The message length of each active \ac{NU} is $421$ bits. The message is encoded by a $(511,421,10)$ BCH code, a null bit is appended to the codeword, and finally, the encoded bits are mapped onto a QPSK constellation, yielding a data payload of $\Nd=256$ symbols.
The payload is appended to a pilot sequence of $\Np=8$ symbols to form the packet. 
Each \ac{NU} is configured to send exactly $R$ replicas of its packet towards the \ac{AP} \ac{ELAA}. In this configuration, each \ac{NU} has up to $R_\text{max}=20$ orthogonal transmission directions, leveraging near-field spatial diversity. However, directions resulting in transmission beyond the \ac{ELAA} boundaries are excluded, leaving each \ac{NU} with $R_\text{max}=13$ usable orthogonal directions for packet replication. The clustering threshold parameter introduced in Sec.~\ref{sec:CRA} is set equal to $\eta=2\sigma/\Np$, ensuring robust initial signal detection and reliable performance in the proposed processing scheme.
Monte Carlo simulations were conducted to evaluate the performance of the \ac{CSRA} scheme in terms of \ac{PLR}, i.e., $\PL$, as a function of the number of simultaneously active devices, $K$, and the number of transmitted packet replicas, $R$.

\subsection{Impact of the Number of Transmitted Replicas} 
\begin{figure}[t]
        \centering
        \resizebox{0.99\columnwidth}{!}{
            \definecolor{aloha1}{rgb}{0.89, 0.82, 0.04}
\definecolor{baseline}{rgb}{0.00000,0.44700,0.74100}%
\definecolor{genie}{rgb}{0.85000,0.32500,0.09800}%
\definecolor{radius30}{rgb}{0.59, 0.29, 0.0}%
\definecolor{radius40}{rgb}{0.0, 0.5, 0.0}%
\definecolor{radius80}{rgb}{0.0, 0.0, 0.0}%
\definecolor{corr2}{rgb}{1.0, 0.49, 0.0}%
\definecolor{corr1}{rgb}{0.6, 0.4, 0.8}
%\definecolor{amber}
%
\begin{tikzpicture}

\begin{axis}[%
scale only axis,
xmin=25,
xmax=50,
xtick distance = 5,
xlabel style={font=\large},
xlabel={$K$},
xlabel shift=-5pt,
ticklabel style = {font=\large},
ymode=log,
ymin=1e-4,
ymax=1,
yminorticks=true,
ylabel style={font=\large},
ylabel={$P_{\mathrm{L}}$},
axis background/.style={fill=white},
xmajorgrids,
ymajorgrids,
yminorgrids,
legend style={at={(0.01,0.99)}, anchor=north west, legend cell align=left, align=left, draw=white!15!black, fill opacity=0.8},
legend entries={%
SIC, % linea continua
No SIC, % linea tratteggiata
$R=2$,
$R=3$,
$R=4$,
$R=5$,
}
]

% Linea continua per SIC
\addlegendimage{line legend, color=gray, line width=1.6pt}
% Linea tratteggiata per no SIC
\addlegendimage{dashed, color=gray, line width=1.6pt}

% Marker per i valori di R
\addlegendimage{only marks, line width=1.4pt, mark=diamond, color=radius80, mark size=3.5pt,  mark options={solid, radius80}}
\addlegendimage{only marks, line width=1.4pt, mark=triangle, color=baseline, mark size=3.5pt, mark options={solid, baseline}}
\addlegendimage{only marks, line width=1.4pt, mark=o, color=corr1, mark size=3.2pt, mark options={solid, corr1}}
\addlegendimage{only marks, line width=1.4pt, mark=square, color=baseline, mark size=3.2pt, mark options={solid, genie}}

\addplot [color=radius80, dashed, line width=1.8pt, mark size=3.5pt, mark=diamond, mark options={solid, radius80}, clip mode=individual]
  table[row sep=crcr]{%
20 0.001299\\
25 0.022753\\
30 0.044010\\
35 0.073451\\
40 0.097304\\
45 0.117386\\
50 0.149542\\
};

%\addlegendentry{$R=2$, no SIC}

\addplot [color=radius80, line width=1.8pt, mark size=3.5pt, mark=diamond, mark options={solid, radius80}, clip mode=individual]
  table[row sep=crcr]{%
20 0.000388\\
25 0.002468\\
30 0.002995\\
35 0.003349\\
40 0.004572\\
45 0.006013\\
50 0.010791\\
};

%\addlegendentry{$R=2$, SIC}

\addplot [color=baseline, dashed, line width=1.8pt, mark size=3.5pt, mark=triangle, mark options={solid, baseline}, clip mode=individual]
  table[row sep=crcr]{%
20 0.011359\\
25 0.022795\\
30 0.043717\\
40 0.112182\\
50 0.198684\\
};

%\addlegendentry{$R=3$, no SIC}

\addplot [color=baseline, line width=1.8pt, mark size=3.5pt, mark=triangle, mark options={solid, baseline}, clip mode=individual]
  table[row sep=crcr]{%
20 0.000388\\
25 0.000486\\
30 0.000534\\
40 0.001059\\
50 0.003316\\
};

%\addlegendentry{$R=3$, SIC}

\addplot [color=corr1, dashed, line width=1.8pt, mark size=3.2pt, mark=o, mark options={solid, corr1}, clip mode=individual]
  table[row sep=crcr]{%
20 0.011897\\
25 0.028679\\
30 0.058775 \\
35 0.103484\\
40 0.158102\\
45 0.227467\\
50 0.307678\\
};
%\addlegendentry{$R=4$, no SIC}

\addplot [color=corr1, line width=1.8pt, mark size=3.2pt, mark=o, mark options={solid, corr1}, clip mode=individual]
  table[row sep=crcr]{%
20 0.000000\\
25 0.000129\\
30 0.000203\\
35 0.000307\\
40 0.000946\\
45 0.003168\\
50 0.010328\\
};
%\addlegendentry{$R=4$, SIC}

\addplot [color=genie, dashed, line width=1.8pt, mark size=3.2pt, mark=square, mark options={solid, genie}, clip mode=individual]
  table[row sep=crcr]{%
20 0.017617\\
25 0.041985\\
30 0.085152\\
35 0.155338\\
40 0.242253\\
45 0.344852\\
50 0.480241\\
};
%\addlegendentry{$R=5$, no SIC}

\addplot [color=genie, line width=1.8pt, mark size=3.2pt, mark=square, mark options={solid, genie}, clip mode=individual]
  table[row sep=crcr]{%
20 0.000000\\
25 0.000153\\
30 0.000303\\
35 0.000538\\
40 0.002774\\
45 0.014203\\
50 0.085372\\
};
%\addlegendentry{$R=5$, SIC}

\end{axis}
\end{tikzpicture}%
        }
        \vspace{-.7cm}\caption{Packet loss rate $\PL$ of the proposed CSRA scheme as a function of the number of active users $K$, with and without SIC.}
        \label{fig:PL_vs_K_replicas}
\end{figure}
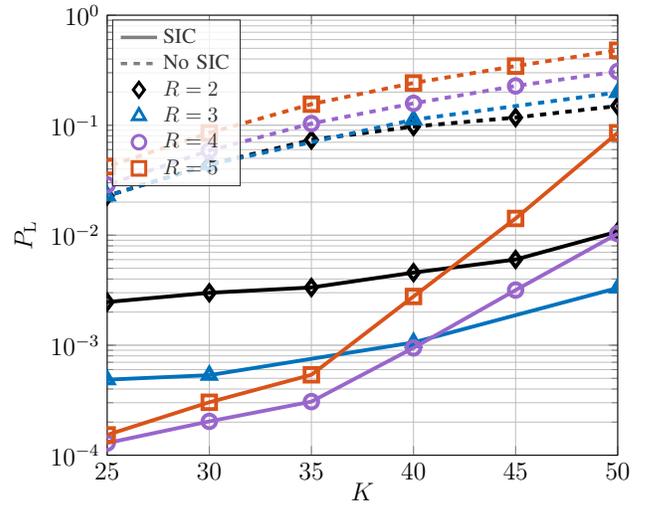

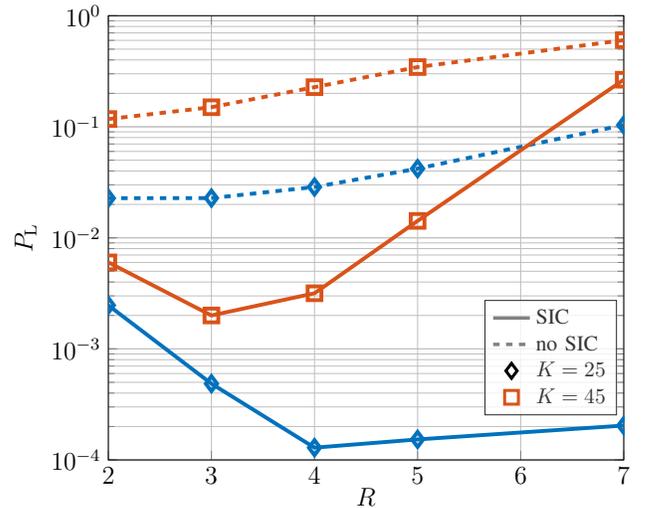
\begin{figure}[t]
        \centering
        \resizebox{0.99\columnwidth}{!}{
            \definecolor{aloha1}{rgb}{0.89, 0.82, 0.04}
\definecolor{baseline}{rgb}{0.00000,0.44700,0.74100}%
\definecolor{genie}{rgb}{0.85000,0.32500,0.09800}%
\definecolor{radius30}{rgb}{0.59, 0.29, 0.0}%
\definecolor{radius40}{rgb}{0.0, 0.5, 0.0}%
\definecolor{radius80}{rgb}{0.0, 0.0, 0.0}%
\definecolor{corr2}{rgb}{1.0, 0.49, 0.0}%
\definecolor{corr1}{rgb}{0.6, 0.4, 0.8}
%\definecolor{amber}
%
\begin{tikzpicture}

\begin{axis}[%
scale only axis,
xmin=2,
xmax=7,
xtick distance = 1,
xlabel style={font=\large},
xlabel={$R$},
xlabel shift=-5pt,
ticklabel style = {font=\large},
ymode=log,
ymin=1e-4,
ymax=1,
yminorticks=true,
ylabel style={font=\large},
ylabel={$P_{\mathrm{L}}$},
axis background/.style={fill=white},
xmajorgrids,
ymajorgrids,
yminorgrids,
legend style={at={(0.73,0.36)}, anchor=north west, legend cell align=left, align=left, draw=white!15!black, fill opacity=0.8},
legend entries={%
SIC, % linea continua
no SIC, % linea tratteggiata
{$K=25$},
{$K=45$},
}
]

% Linea continua per SIC
\addlegendimage{line legend, color=gray, line width=1.6pt}
% Linea tratteggiata per no SIC
\addlegendimage{dashed, color=gray, line width=1.6pt}

% Marker per i valori di R
\addlegendimage{only marks, line width=1.4pt, mark=diamond, color=radius80, mark size=3.5pt,  mark options={solid, radius80}}
\addlegendimage{only marks, line width=1.4pt, mark=square, color=genie, mark size=3.2pt, mark options={solid, genie}}

\addplot [color=baseline, dashed, line width=1.8pt, mark size=3.5pt, mark=diamond, mark options={solid, baseline}, clip mode=individual]
  table[row sep=crcr]{%
2 0.022753\\
3 0.022795\\
4 0.028679\\
5 0.041985\\
7 0.103803\\
};

%\addlegendentry{$K=25$, no SIC}

\addplot [color=baseline, line width=1.8pt, mark size=3.5pt, mark=diamond, mark options={solid, baseline}, clip mode=individual]
  table[row sep=crcr]{%
2 0.002468\\
3 0.000486\\
4 0.000129\\
5 0.000153\\
7 0.000204\\
};

%\addlegendentry{$K=25$, SIC}

\addplot [color=genie, dashed, line width=1.8pt, mark size=3.2pt, mark=square, mark options={solid, genie}, clip mode=individual]
  table[row sep=crcr]{%
2 0.117386\\
3 0.15\\
4 0.227467\\
5 0.344852\\
7 0.602442\\
};
%\addlegendentry{$K=45$, no SIC}

\addplot [color=genie, line width=1.8pt, mark size=3.2pt, mark=square, mark options={solid, genie}, clip mode=individual]
  table[row sep=crcr]{%
2 0.006013\\
3 0.002\\
4 0.003168\\
5 0.014203\\
7 0.265325\\
};
%\addlegendentry{$K=45$, SIC}

\end{axis}
\end{tikzpicture}%
        }
        \vspace{-.7cm}\caption{Packet loss rate $\PL$ of the proposed CSRA scheme as a function of the number of replicas $R$, with and without SIC.}
        \label{fig:PL_vs_R}
\end{figure}

Fig.~\ref{fig:PL_vs_K_replicas} shows the \ac{PLR} of the proposed \ac{CSRA} scheme, both with and without \ac{SIC}, as a function of the number of active \acp{NU} for various numbers of transmitted packet replicas, $R$. The results demonstrate the essential role of \ac{SIC} in ensuring the reliability of the scheme. Without employing \ac{SIC}, the system faces challenges in reaching even a packet loss probability of $\PL=10^{-2}$ in the given scenario.
The results also highlight that an optimal choice of the number of transmitted packet replicas, $R$, depends on the system load. This is also evident in Fig.~\ref{fig:PL_vs_R}, which shows the \ac{PLR} of \ac{CSRA} scheme varying $R$ for $K=25$ and $K=45$ active devices. When the number of active users is relatively low, e.g., around $K=25$, it is preferable to set $R=4$, which enables the system to achieve a $\PL$ close to $10^{-4}$. However, as the traffic load increases, transmitting fewer replicas reduces interference and improves overall performance. For instance, with $K=45$ active users setting $R=5$ yields a $\PL$ around $3\cdot 10^{ 
-3}$, while $R=4$ further reduces $\PL$ to approximately $2\cdot10^{-3}$.
These results suggest a dynamic configuration of $R$ based on traffic load, balancing the benefit of additional replicas with the resulting interference. Transmitting more replicas boosts reliability in low-load scenarios but introduces more interference under high-load conditions, thus impacting $\PL$.

\subsection{Comparison with CSRA-SE}
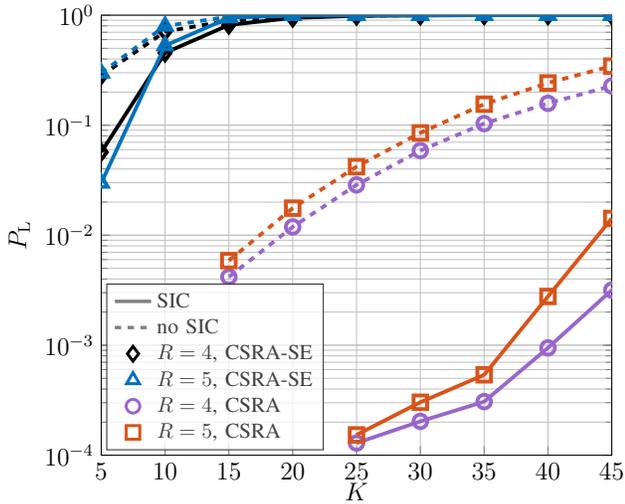
\begin{figure}[t]
        \centering
        \resizebox{0.99\columnwidth}{!}{
            \definecolor{aloha1}{rgb}{0.89, 0.82, 0.04}
\definecolor{baseline}{rgb}{0.00000,0.44700,0.74100}%
\definecolor{genie}{rgb}{0.85000,0.32500,0.09800}%
\definecolor{radius30}{rgb}{0.59, 0.29, 0.0}%
\definecolor{radius40}{rgb}{0.0, 0.5, 0.0}%
\definecolor{radius80}{rgb}{0.0, 0.0, 0.0}%
\definecolor{corr2}{rgb}{1.0, 0.49, 0.0}%
\definecolor{corr1}{rgb}{0.6, 0.4, 0.8}
%\definecolor{amber}
%
\begin{tikzpicture}

\begin{axis}[%
scale only axis,
xmin=5,
xmax=45,
xtick distance = 5,
xlabel style={font=\large},
xlabel={$K$},
xlabel shift=-5pt,
ticklabel style = {font=\large},
ymode=log,
ymin=1e-4,
ymax=1,
yminorticks=true,
ylabel style={font=\large},
ylabel={$P_{\mathrm{L}}$},
axis background/.style={fill=white},
xmajorgrids,
ymajorgrids,
yminorgrids,
legend style={at={(0.01,0.39)}, anchor=north west, legend cell align=left, align=left, draw=white!15!black, fill opacity=0.75},
legend columns=1,
legend entries={%
SIC, % linea continua
no SIC, % linea tratteggiata
{$R=4$, CSRA-SE},
{$R=5$, CSRA-SE},
{$R=4$, CSRA},
{$R=5$, CSRA},
}
]

% Linea continua per SIC
\addlegendimage{line legend, color=gray, line width=1.6pt}
% Linea tratteggiata per no SIC
\addlegendimage{dashed, color=gray, line width=1.6pt}

% Marker per i valori di R
\addlegendimage{only marks, line width=1.4pt, mark=diamond, color=radius80, mark size=3.5pt,  mark options={solid, radius80}}
\addlegendimage{only marks, line width=1.4pt, mark=triangle, color=baseline, mark size=3.5pt, mark options={solid, baseline}}
\addlegendimage{only marks, line width=1.4pt, mark=o, color=corr1, mark size=3.2pt, mark options={solid, corr1}}
\addlegendimage{only marks, line width=1.4pt, mark=square, color=baseline, mark size=3.2pt, mark options={solid, genie}}

\addplot [color=radius80, dashed, line width=1.8pt, mark size=3.5pt, mark=diamond, mark options={solid, radius80}, clip mode=individual]
  table[row sep=crcr]{%
5 0.282975\\
10 0.709241\\
15 0.892358\\
20 0.962416\\
25 0.991475\\
30 0.999245\\
35 1\\
40 1\\
45 1\\
50 1\\
};

%\addlegendentry{$R=4$, SE, no SIC}

\addplot [color=radius80, line width=1.8pt, mark size=3.5pt, mark=diamond, mark options={solid, radius80}, clip mode=individual]
  table[row sep=crcr]{%
5 0.056860\\
10 0.454455\\
15 0.815285\\
20 0.944966\\
25 0.986885\\
30 0.994175\\
35 1\\
40 1\\
45 1\\
50 1\\
};

%\addlegendentry{$R=4$, SE, SIC}

\addplot [color=baseline, dashed, line width=1.8pt, mark size=3.5pt, mark=triangle, mark options={solid, baseline}, clip mode=individual]
  table[row sep=crcr]{%
5 0.297872\\
10 0.797143\\
15 0.967273\\
20 0.996429\\
25 1.000000\\
30 1\\
35 1\\
40 1\\
45 1\\
50 1\\
};

%\addlegendentry{$R=5$, SE, no SIC}

\addplot [color=baseline, line width=1.8pt, mark size=3.5pt, mark=triangle, mark options={solid, baseline}, clip mode=individual]
  table[row sep=crcr]{%
5 0.029787\\
10 0.528571\\
15 0.946667\\
20 0.994048\\
25 1.000000\\
30 1\\
35 1\\
40 1\\
45 1\\
50 1\\
};

%\addlegendentry{$R=5$, SE, SIC}

\addplot [color=corr1, dashed, line width=1.8pt, mark size=3.2pt, mark=o, mark options={solid, corr1}, clip mode=individual]
  table[row sep=crcr]{%
15 0.004173\\
20 0.011897\\
25 0.028679\\
30 0.058775 \\
35 0.103484\\
40 0.158102\\
45 0.227467\\
50 0.307678\\
};
%\addlegendentry{$R=4$, no SIC}

\addplot [color=corr1, line width=1.8pt, mark size=3.2pt, mark=o, mark options={solid, corr1}, clip mode=individual]
  table[row sep=crcr]{%
20 0.000000\\
25 0.000129\\
30 0.000203\\
35 0.000307\\
40 0.000946\\
45 0.003168\\
50 0.010328\\
};
%\addlegendentry{$R=4$, SIC}

\addplot [color=genie, dashed, line width=1.8pt, mark size=3.2pt, mark=square, mark options={solid, genie}, clip mode=individual]
  table[row sep=crcr]{%
15 0.005873\\
20 0.017617\\
25 0.041985\\
30 0.085152\\
35 0.155338\\
40 0.242253\\
45 0.344852\\
};
%\addlegendentry{$R=5$, no SIC}

\addplot [color=genie, line width=1.8pt, mark size=3.2pt, mark=square, mark options={solid, genie}, clip mode=individual]
  table[row sep=crcr]{%
20 0.000000\\
25 0.000153\\
30 0.000303\\
35 0.000538\\
40 0.002774\\
45 0.014203\\
};
%\addlegendentry{$R=5$, SIC}

\end{axis}
\end{tikzpicture}%
        }
        \vspace{-.7cm}\caption{Packet loss rate $\PL$ of the proposed CSRA scheme as a function of the number of active users $K$, with and without SIC. Comparison with single-element (SE) processing at the ELAA.}
        \label{fig:CSRA-SE}
\end{figure}

Since the grant-free nature of this communication setup prevents the \ac{AP} from performing a priori channel estimation or identifying active nodes, comparisons with grant-based methods are unsuitable. Instead, we benchmark the performance against a single-element variant of the \ac{CSRA} scheme, the CSRA-SE one. In CSRA-SE, detection and decoding are performed on individual antenna elements independently, without clustering, which would otherwise enable linear combining of received signals.
Fig.~\ref{fig:CSRA-SE} compares the \ac{CSRA} and CSRA-SE schemes in terms of $\PL$ varying $K$, for different numbers of transmitted replicas $R$, both with and without \ac{SIC}. The results indicate that processing signals independently at each \ac{AP} antenna element, as in CSRA-SE, leads to significantly lower performance. 
The CSRA-SE scheme is unable to reach $\PL=10^{-2}$ even with only $K=5$ active users. This drop in performance is mainly due to the system operating in a low \acl{SNR} regime, where sufficient performance cannot be attained without utilizing the array gain enabled by clustering and combining techniques at the receiver.

\section{Conclusions}\label{sec:Conclusions}
This work introduced a novel grant-free coded random access scheme tailored to meet the demands of massive \ac{MTC} in \ac{IIoT} networks. By leveraging an \ac{ELAA} at the \ac{AP}, and exploiting spatial multiplexing within the near field region, the proposed approach effectively mitigates interference and enhances reliability, even in dense, uncoordinated \ac{IIoT} environments.
Specifically, \ac{SIC} is exploited with clustering of the signals received at multiple \ac{AP} antennas, thus circumventing the need for extensive \ac{CSI} estimation and effectively leveraging the beamforming gain. Extensive simulations demonstrate that the \ac{CSRA} scheme significantly reduces the \ac{PLR} compared to single-element processing, which cannot leverage the beamforming gain enabled by clustering. 
Furthermore, optimal performance is achieved by dynamically adjusting the number of transmitted replicas based on network load, balancing interference and reliability. This adaptability makes \ac{CSRA} a robust and scalable solution for next-generation \ac{IIoT} applications.

\section*{Acknowledgments}
This work was partially supported by the European Union under the Italian National Recovery and Resilience Plan (NRRP) of NextGenerationEU, partnership on ``Telecommunications of the Future" (PE00000001 - program ``RESTART"), and in part by the HORIZON-JU-SNS-2022-STREAM-B-01-03 6G-SHINE Project under Grant 101095738. Giulia Torcolacci was funded by an NRRP Ph.D. grant.

\bibliographystyle{IEEEtran}
\bibliography{IEEEabrv,references}

\end{document}